\documentclass[aps,nofootinbib,showpacs,%twocolumn,
floatfix,superscriptaddress]{revtex4}
\usepackage{graphicx}
\usepackage{epsf,amsfonts,amssymb,amsbsy}
\usepackage[mathscr]{eucal}

\begin{document}

\title{Quasiattractor in models of new and chaotic inflation}
\author{V.V.Kiselev}
\email{Valery.Kiselev@ihep.ru}
 \affiliation{Russian State Research
Center ``Institute for High
Energy Physics'', %\\
Pobeda 1, Protvino, Moscow Region, 142281, Russia\\ Fax:
+7-4967-742824}
 \affiliation{Moscow Institute of Physics and
Technology, Institutskii per. 9, Dolgoprudnyi, Moscow Region,
141701, Russia}
\author{S.A.Timofeev}
\affiliation{Moscow Institute of Physics and Technology,
Institutskii per. 9, Dolgoprudnyi, Moscow Region, 141701, Russia}
 \pacs{98.80.-k}
\begin{abstract}
Inflation with a scalar-field potential of the form $\lambda
(\phi^2-v^2)^2$ can be described in terms of a parametrical
attractor with critical points, whose driftage depends on the
control value of the slowly changing Hubble rate. The method allows
us to easily obtain theoretical expressions for fluctuations of
inhomogeneity in both the cosmic microwave background and
distribution of matter. We find the region for admissible values of
potential parameters, wherein theoretical predictions are consistent
with experimental results within the limits of measurement
uncertainties.
\end{abstract}
\maketitle

\section{Introduction}

At present, in cosmology there is a problem in determining the
parameters of inflation, which, in fact, has become the standard
model for the early stage of Universe evolution
\cite{i-Guth,i-Linde,i-Albrecht+Steinhardt,i-Linde2,inflation}
\textit{before} the Big Bang. The Big Bang is considered now as a
short stage of reheating of the Universe due to a transformation of
inflaton energy into the energy of matter, whereas the inflaton is
usually ascribed to be a real scalar field. In this respect it would
be useful to have a complete arsenal of effective methods in order
to describe various characteristics at the inflationary stage. These
instruments would allow us to carry out a more thorough analysis of
theoretical models in comparison with quite precise modern
experimental data. At present, the basic tool of such studies is the
slow-roll approximation in the field equations of inflation (see the
review in \cite{inflation}).

The slow-roll dynamics of evolution can be consistently treated in
the framework of a $1/N$-expansion at a large amount of e-folding
$N$ for the scale factor of expansion, which was presented in
\cite{BdeVegaS1} as a general analysis of relative scaling behavior
of inflaton quantities versus $1/N$. So, the inflaton potential $V$
gets the characteristic scale $M$ by $V\sim N\,M^4$ at $M\sim
10^{16}$ GeV, while the inflaton field $\phi$ behaves like
$\phi\sim\sqrt{N}\,M_\mathrm{Pl}$ with $M_\mathrm{Pl}$ being the
Planck mass as given by the Newton gravitational constant
$G=1/M^2_\mathrm{Pl}$. In this respect, one could expect, for
instance, a characteristic value of quartic coupling in the inflaton
self-action of the order of $\lambda\sim
1/N\,\big(M/M_\mathrm{Pl}\big)^4\sim 10^{-14}$. Thus, one gets the
tool of strict consideration of the slow-rolling regime.

We follow another way, which is the method of the quasiattractor.
This approach was offered in \cite{Mexicans} for the case of a
quadratic potential, in order to generalize and develop
investigations considering the dependence of cosmological evolution
on initial data that were performed in
\cite{BGKhZ,PW,KLS,laM-P,KBranden}. Further, we have applied the
same approach to the potential of $\lambda\phi^4$ in \cite{KT3}.
This kind of potential refers to the models of ``chaotic
inflation'', when the evolution occurs from large fields at Planck
scales towards the global minimum at $\phi=0$. However, it would be
useful to somehow generalize these results to a potential of the
form $\lambda (\phi^2-v^2)^2$, permitting, first, the opportunity of
a situation with the scenario of ``new inflation'', when the field
evolves from a position in the vicinity of a local maximum at zero
value of the field to the global minimum at $\phi=v$. Second, as we
will see, such a potential allows us to essentially expand the
region of admissible values of the potential parameters consistent
with the data. This fact significantly increases the viability of
the model.

The quasiattractor approach can be described by the following: After
choosing the model potential we derive the equations of the system
motion, which are generally not analytically soluble, so that we try
to treat the problem by applying some consistent approximations in
order to describe the system evolution. In the method of the
quasiattractor we introduce new dimensionless variables with
presumed properties of scaling. Then, the differential equations of
the first order can be considered as an autonomous system. The
system could attain stable critical points on a phase plane. The
trajectories converge to these points, being the attractors. Our
first task is to search for such critical points. The notion of
"quasiattractor"  refers to the stable critical point of an
autonomous system\footnote{The exact attractor arises at quite
definite fixed functional forms of potential: at zero cosmological
constant it is the exponent \cite{Wetterich,CLW,FJ,Alb_Skordis},
while at nonzero cosmological constant it is the hyperbolic cosine
\cite{Kiselev-attract}. } with external parameters slowly drifting
with the evolution. The position of the critical point is not fixed,
since it is determined by the control parameters, which evolve and
displace the critical point. But the evolution velocity of the
control parameters is slow enough in order to consider the
displacement of the point in the phase space as driftage. Thus, the
system motion is the following: The system very quickly ``falls'' to
the quasiattractor in the phase space, i.e. to the stable critical
point slowly drifting during the evolution. So, the information
about the initial position of the system is lost, while values of
control parameters determining the position of attractor are
important. The further evolution of system is exclusively determined
by the driftage of the quasiattractor. The system motion is
appropriated by the evolution of control parameters, and the system
seems to lose some degrees of freedom.

As we will derive below, the driftage of the attractor is equivalent
to the slow-roll regime of inflation treated in the framework of
$1/N$-expansion considered in \cite{BdeVegaS1,BdeVegaS2}.

In Section  \ref{II} we consider mathematical aspects of the
quasiattractor, while in Section \ref{III} we compare theoretical
results with experimental data. Our results are in agreement with
the precise analysis of a complete data set previously performed in
\cite{DdeVegaS1,DdeVegaS2} in the framework of Monte Carlo Markov
Chains. In the Conclusion we discuss the results obtained.

\section{Mathematical Aspects\label{II}}

\subsection{Equations of Motion}

Let us consider the action of the inflaton in the form
\begin{equation}
    S=\int dx^4 \sqrt{-g} \left\{\frac{1}{2}\,\partial_\mu\phi \,\partial^\mu\phi -
    V(\phi)\right\}
\end{equation}
with the potential
\begin{equation}
V=\frac{\lambda}{4}(\phi^2-v^2)^2.
\end{equation}
The evolution of a homogeneous isotropic flat Universe is described
by a Friedmann-Lemaitre-Robertson-Walker metric (FLRW) in Cartesian
coordinates
\begin{equation}
g_{\mu \nu}=\mbox{diag}(1,-a^2(t),-a^2(t),-a^2(t)),
\end{equation}
where $a(t)$ is the scale factor with its usual physical
interpretation. The evolution equations read off as
\begin{eqnarray}
 \ddot\phi &=&-3H\dot\phi-\lambda\phi(\phi^2-v^2),
 \label{4}\\[1mm]
 \dot H &=& -4\pi G \dot\phi^2.\label{5}
\end{eqnarray}
Here the dot denotes the derivative with respect to time $t$. The
Hubble rate is defined by $H=\dot a/a$.

The Friedmann relation is derived from (\ref{4}) and (\ref{5}), so that
\begin{equation}\label{6}
H^2=\frac{4\pi G}{3}\left\{\dot
\phi^2+\frac{1}{2}\lambda(\phi^2-v^2)^2\right\}.
\end{equation}
All of the equations (\ref{4})--(\ref{6}) are consequences of
General Relativity.

\subsection{Autonomous System}

For the sake of simplicity let us change variables:
\begin{eqnarray}
x&=&\frac{\kappa}{\sqrt{6}}\,\frac{\dot \phi}{H}, \\
y&=&\sqrt[4]{\frac{\lambda}{12}}\frac{\sqrt{\kappa}}{\sqrt{H}}
\sqrt{|\phi^2-v^2|},\label{8} \\
z&=&\frac{\sqrt[4]{3\lambda}}{\sqrt{\kappa H}}, \\
u&=&\frac{\kappa v}{\sqrt{6}},
\end{eqnarray}
where $\kappa^2=8 \pi G$. Then, Eqs. (\ref{4})--(\ref{6}) take the
form\footnote{For brevity of notation, we take square roots in the
arithmetic sense that corresponds to the case when the field takes
values greater than the vacuum expectation, $|\phi| > v$, so that
the scenario of chaotic inflation is realized. Otherwise, in the
case of new inflation with $|\phi| < v$, one should take the root
with the opposite sign under the substitution $y^2 \to -y^2$ in the
radicand. This procedure is equivalent to removing the absolute
value of the radicand in expression (\ref{8}), so that $y^2$ can
formally run to negative values in the model of new inflation.}
\begin{eqnarray}
x^\prime&=&3x^3-3x-2y^2z\sqrt{y^2+u^2z^2},\label{11}\\
yy^\prime&=&\frac{3}{2}x^2y^2+xz\sqrt{y^2+u^2z^2},\label{12}\\
z^\prime&=&\frac{3}{2}x^2z \label{N_z},\label{13}
\end{eqnarray}
where the prime denotes the derivative with respect to
$N=\ln{(a/a_\mathrm{init.})}$. Then the relation
$\frac{\partial}{\partial t}=H \frac{\partial}{\partial N}$ is
valid. The physical sense of $N$ is that it counts the amount of
e-folding during the expansion of the Universe from
$t_\mathrm{init.}$ till the current point, i.e. when the scale
factor increases by $e^N$ times.

In terms of the new variables the Friedmann relation reads off as
\begin{equation}\label{Fr2}
x^2+y^4=1.
\end{equation}
The equations are simplified, since they are already differential
equations of the first order, though they are nonhomogeneous, which
are easier for analysis than the initial ones.

Indeed, the two equations of (\ref{11}) and (\ref{12}) can be
considered as an autonomous system of differential equations of the
first order with external parameter $z$. Then, there is a question
of the stability of given system. The numerical analysis shows, that
the system is stable under some definite conditions. The control
parameter of autonomous system is the slowly varying quantity $z$.
Obviously, the driftage proceeds smoothly at $x^2 z\ll 1$. The
question is when will the critical point be stable? Then, all of
trajectories will approach this point, and it becomes the
parametrical attractor, i.e. the quasiattractor, while the system,
gradually having come to it, will remain at the critical point and
drift together with it, and the evolution of the actually stable
point (the quasiattractor) will be determined by the control
quantity $z$.

Equations for the quasiattractor ($x^\prime=y^\prime=0$) are the
following:
\begin{eqnarray}
3x^3-3x-2y^2z\sqrt{y^2+u^2z^2}=0,\label{15}\\
\frac{3}{2}xy^2+z\sqrt{y^2+u^2z^2}=0. \label{16}
\end{eqnarray}
It is worth noticing that the system of equations is compatible with
the Friedmann condition.

For the sake of simplification of system, we can make the
following change of variable:
\begin{equation}
Y^2=y^2+u^2z^2.
\end{equation}
Then, the equations transform and look less cumbersome without
radicals:
\begin{eqnarray}
x^\prime&=&3x^3-3x-2Yz(Y^2-u^2z^2),\\
Y^\prime&=&\frac{3}{2}x^2Y+xz,\\
z^\prime&=&\frac{3}{2}x^2z,\\
1&=&x^2+(Y^2-u^2z^2)^2.
\end{eqnarray}
The scaling properties of $Y$ and $y$ are equivalent, since these
quantities differ by a shift, which depends on the external
parameter controlling the driftage.

If the variable $y$ is eliminated from the system, the equations for
the critical point in the physical case of $x\neq 0$, $y\neq 0$ are
reduced to the single equation in $x$ (we recall that the quantity
$z$ is the parameter)
\begin{equation}
\frac{3}{2}x\sqrt{1-x^2}+z\sqrt{\sqrt{1-x^2}+u^2z^2}=0. \label{x}
\end{equation}

\subsection{Analysis of the System}

Let us analyze the stability of the critical point $(x_c,\,y_c)$.
Introduce small deviations from the attractor $(\delta x,\,\delta
y)$, then $x=x_c+\delta x,\; y=y_c+\delta y$. We obtain the
following differential equations for the deviations in the linear
approximation:
\begin{eqnarray}
\left(
\begin{array}{c}
\delta x^\prime \\ \delta y^\prime
\end{array}
\right) = \left(
\begin{array}{ccc}
9x^2_c-3 & \displaystyle\frac{4z^2(3y_c^2+2u^2z^2)}{3x_cy_c}
\\[5mm] \displaystyle
\frac{3}{2}\,x_cy_c &
\displaystyle\frac{9x_c^2y_c^4+4u^2z^4}{6y_c^4}
\end{array}
\right) \left(
\begin{array}{c}
\delta x \\ \delta y
\end{array}
\right).
\end{eqnarray}
The Friedmann condition requires
\begin{equation}
x_c\delta x+2y^3_c\delta y=0,
\end{equation}
or
\begin{equation}
\left(
\begin{array}{ccc}
x_c & 2y_c^3
\end{array}
\right) \left(
\begin{array}{ccc}
\delta x \\ \delta y
\end{array}
\right)=0,
\end{equation}
i.e. the solution, satisfying the Friedmann equation, will be
proportional to the eigenvector
\begin{equation}
v \sim \left(
\begin{array}{ccc}
2y^3_c \\-x_c
\end{array}
\right).
\end{equation}
Such an eigenvector for the given matrix exists, and is single with
the eigenvalue
\begin{equation}
\mathscr{B}=3-6y^4-\frac{2}{3}\,\frac{z^2}{y^2}.
\end{equation}
Therefore, the evolution goes according to the law
\begin{equation}
\left(
\begin{array}{ccc}
\delta x \\ \delta y
\end{array}
\right)=C\left(
\begin{array}{ccc}
2y^3_c \\-x_c
\end{array}
\right) e^{\mathscr{B} N}.
\end{equation}
Thus, we see, that this is the only solution of system. It satisfies
the imposed constraints. Further advancement of the analysis of the
autonomous system will consist in the direct examination of the
stability of the obtained solution.

We require $\mathscr{B} <0$ for stability of attractor. This
condition is valid at small values of $x$ and $z$ (then, according
to the Friedmann equation $y^4$ is close to 1) and $\mathscr{B}$ is
certainly less than zero. The constraint on the smallness of $x$ and
$z$ is actually valid, as we will see below.

\subsection{The Universe Inflation}

Let us consider Universe inflation due to the inflaton with the
chosen potential. The condition of accelerated expansion is the
following:
\begin{equation}
\ddot{a}>0 \quad\Rightarrow \quad
\frac{\ddot{a}}{a}=\dot{H}+H^2>0\quad \Rightarrow\quad
\frac{\dot{H}}{H^2}=-3x^2>-1.
\end{equation}
Accordingly, such an expansion regime ends up with
\begin{eqnarray}
x^2_\mathrm{end}&=&\frac{1}{3},\\
y^4_\mathrm{end}&=&\frac{2}{3},\\
z_\mathrm{end}^2&=&\frac{\sqrt{3u^2+1}-1}{u^2\sqrt{6}}.
\end{eqnarray}

During the actual process of expansion with acceleration, the
quantities should satisfy the inequalities,
\begin{eqnarray}
x^2_c&<&x^2_\mathrm{end},\\
y^4_c&>&y^4_\mathrm{end},\label{34}\\
z^2&<&z^2_\mathrm{end}.
\end{eqnarray}
One can see that such values are in agreement with the condition
making the attractor stable ($\mathscr{B} <0$), since (\ref{34})
gives
$$
    \mathscr{B}<-1-\frac{2}{3}\frac{z^2}{y^2}.
$$
hence, the accelerated expansion is governed by the stable
quasiattractor.

\subsection{Characteristics of The Universe Expansion}

First of all, we determine how many times the Universe expands from
the initial state marked by ``in'' to the end of inflation marked by
``end''. The total amount of e-folding $N$ is given by following
expression, which follows from the equation for the parameter $z$ in
(\ref{N_z}),
\begin{equation}
N_\mathrm{total}=\frac{2}{3}\int^{z_\mathrm{end}}_{z_\mathrm{in}}\frac{dz}{x^2_cz}.
\label{N_total}
\end{equation}
Here, the parameter $x$ is set at the point of the quasiattractor.

In order to find numerical values of the theory parameters, one
should compare it with observational data.  Experiment measures the
inhomogeneity of the cosmic microwave background, related to the
inhomogeneity of matter, also independently measured, hence we need
to find the distribution of the inflaton inhomogeneity, which leads
to the matter inhomogeneity at the stage of reheating. Such
inhomogeneity is given by the quantum fluctuations of the inflaton.
Then, the spectral density of scalar and tensor perturbations look
as
\begin{equation}
P_S(k)=\left(\frac{H}{2\pi}\right)^2\left(\frac{H}{\dot{\phi}}\right)^2
=\frac{\lambda}{8\pi^2}\,\frac{1}{x_c^2z^4}
\end{equation}
and
\begin{equation}
P_T(k)=8\kappa^2\left(\frac{H}{2\pi}\right)^2
=\frac{6\lambda}{\pi^2}\,\frac{1}{z^4},
\end{equation}
where the wave vector $k$ is determined by the Hubble rate at the
exit of fluctuations from the horizon, i.e. at $k=aH$.

Consider the ratio $r$ determining the relative contribution of
tensor spectrum,
\begin{equation}
r=\frac{P_T(k)}{P_S(k)}=48\,x_c^2,
\end{equation}
and define the spectral index $n_S$ as
\begin{equation}
n_S-1\equiv \frac{d\ln P_S}{d\ln k}.
\end{equation}
One can easily see that
\begin{equation}
\ln\frac{k}{k_\mathrm{end}}=N-2\ln\frac{z}{z_\mathrm{end}},
\end{equation}
so that differentiation with respect to the wave vector is reduced
to derivative with respect to the parameter $z$, determining the
dynamics in the method of the quasiattractor.

\subsection{Finding the Total $N$}

From equation (\ref{x}) at $x^2\ll 1$ one approximately
gets\footnote{Let us recall, that (\ref{x_z}) is valid for chaotic
inflation, while in the scenario of new inflation one should change
the sign of $y^2$, i.e. one puts $x_c^2 \approx \frac{4}{9}
z^2(-1+u^2z^2).$}
\begin{equation}
x_c^2 \approx \frac{4}{9} z^2(1+u^2z^2). \label{x_z}
\end{equation}
The Friedmann condition in the forms of (\ref{6}) and (\ref{Fr2})
strictly holds and yields $y^2\approx 1$ in the limit under
consideration, while the attractor position of (\ref{15}),
(\ref{16}) reduced to (\ref{x_z}) gives the slow-roll equation
$3H\dot \phi+\partial V/\partial\phi=0$.

Then, substituting the above expression into (\ref{N_total}), we
obtain
\begin{eqnarray}
N_\mathrm{total} &\approx& \frac{3}{4}
\left(u^2\ln{\frac{1+u^2z^2}{u^2z^2}}-\frac{1}{z^2}\right)
\bigg\vert_{z_\mathrm{in}}^{z_\mathrm{end}} \nonumber\\[3mm]
&=& \frac{3}{4} \left(u^2
\ln{\frac{\sqrt{3u^2+1}-1+\sqrt{6}}{\sqrt{3u^2+1}-1}}-\frac{u^2\sqrt{6}}{\sqrt{3u^2+1}-1}+
\frac{1}{z^2_\mathrm{in}}-u^2\ln{\frac{1+u^2z_\mathrm{in}^2}{u^2z_\mathrm{in}^2}}
\right).
\end{eqnarray}
To simplify this bulky expression we can separate out the function
obtained by substituting $z_\mathrm{end}$, giving the term,
\begin{equation}
F(u)=\frac{3}{4} \left(u^2
\ln{\frac{\sqrt{3u^2+1}-1+\sqrt{6}}{\sqrt{3u^2+1}-1}}-\frac{u^2\sqrt{6}}{\sqrt{3u^2+1}-1}
\right).
\end{equation}
This function monotonically decreases in the interval ${u\in
[0,+\infty)}$, hence, it is restricted by limits at the borders
\begin{equation}
\frac{3}{4} \leqslant F(u) \leqslant \sqrt{\frac{3}{2}}.
\end{equation}
Since the inhomogeneity of the matter spectrum available for
measurements actually refers to $N$ of the order of 60, one can
neglect the contribution of the upper limit in the integral, i.e.
the value of function $F(u)$, to the leading approximation in $1/N$.
Then, the expression for $N_\mathrm{total}$ is simplified to
\begin{equation}\label{46}
N_\mathrm{total} \approx \frac{3}{4}\left(
\frac{1}{z^2_\mathrm{in}}-u^2\ln{\frac{1+u^2z_\mathrm{in}^2}{u^2z_\mathrm{in}^2}}
\right).
\end{equation}

Now we can express $N_\mathrm{total}$ in terms of the
experimentally measured $r$ and $n_S$. We find
\begin{equation}
n_S-1 = \frac{4(3z^2+4z^4u^2)}{4z^2(1+u^2z^2)-3}
=\frac{4(9x_c^2-z^2)}{3(3x_c^2-1)},
\end{equation}
and express all of other parameters as follows:
\begin{equation}
z_\mathrm{in}^2=\frac{12r-3(r-16)(n_S-1)}{64},
\end{equation}
and
\begin{equation}
x_c^2=\frac{r}{48}.
\end{equation}
Making use of the connection between $x_c^2$ and $z^2$ according to
(\ref{x_z}), we get
\begin{equation}\label{50}
u^2=\frac{64(-3r+(n_S-1)(r-16))}{3(-4r+(n_S-1)(r-16))^2}.
\end{equation}

For the sake of simplicity, introduce the quantity $\chi$
\begin{equation}
\chi=4r-(n_S-1)(r-16),
\end{equation}
satisfying the condition of $\chi\leqslant r$ equivalent to
$u^2\geqslant 0$, because
$$
u^2= \frac{64}{3} \frac{r-\chi}{\chi^2}.
$$
Then, we can easily write down the final expression for
$N_\mathrm{total}$
\begin{equation}
N_\mathrm{total}=\frac{16}{\chi}\left\{ 1-\left(
1-\frac{r}{\chi}\right)\ln{\left( 1-\frac{\chi}{r} \right)}
 \right\}.  \label{N_end}
\end{equation}

The above expression can be ``parametrically solved''. So, we introduce
quantity  $\beta\geqslant 0$ according to
$$
    \chi=r(1-\beta).
$$
Then
\begin{equation}\label{r}
    r=\frac{16}{N}\,\frac{1}{1-\beta}\left\{1+
    \frac{\beta}{1-\beta}\,\ln\beta\right\},
\end{equation}
and
\begin{equation}\label{ns}
    n_S-1=-\frac{1-\beta+\beta\ln\beta}{N(1-\beta)^2-1+\beta-\beta\ln\beta}\,
    (3+\beta).
\end{equation}

The formula (\ref{r}) exactly repeats the expression derived in
\cite{DdeVegaS1,DdeVegaS2} in another notation in the framework of
slow-roll approximation, while (\ref{ns}) can match the result of
\cite{DdeVegaS1,DdeVegaS2}, if one neglects subleading terms in the
denominator of (\ref{ns}) at $1/N\to 0$.

In the limits of $\beta\to 0$ and $\beta\to 1$ we obtain the
reference cases with potentials of form $\lambda\phi^4$ and
$m^2\phi^2$, correspondingly. Indeed, up to corrections of the order
of $1/N$ as has been suggested in deriving (\ref{r}) and (\ref{ns}),
we find
$$
u^2=\frac{64}{3} \frac{\beta}{r(1-\beta)^2} \to 0  \qquad \mbox{at
}  \beta \to 0,
$$
hence, the vanishing of the inflaton vacuum expectation value, i.e.
nullifying the quadratic term in the potential, while
$$
\mbox{at } \beta \to 1 \qquad
z^2=\frac{3}{64}\,|\chi|=\frac{3}{64}\,r\,|1-\beta|\to 0,
$$
that has reduced to zero the quartic term in the potential. In
addition,
 \begin{equation}\label{lim1}
    r=\left\{
    \begin{array}{cc}
      \frac{16}{N}, & \beta=0, \\[3mm]
      \frac{8}{N}, & \beta=1, \\
    \end{array}
    \right.
    \qquad
    n_S-1=\left\{
    \begin{array}{cc}
      -\frac{3}{N}, & \beta=0, \\[3mm]
      -\frac{2}{N}, & \beta=1, \\
    \end{array}
    \right.
\end{equation}
in complete consistency with the consideration of these cases in
other approaches.

Generally, the scaling properties of inflation parameters versus $1/N$
are quite complicated because of additional dependence on variable
$\beta$, which can correlate with the amount of e-folding $N$.
Nevertheless, one can see that at fixed $\beta$, the limit of $1/N\to
0$ gives
$$
    r\sim\frac{1}{N},\quad x^2\sim \frac{1}{N},\quad
    z^2\sim\frac{1}{N},\quad u^2\sim{N},\quad
$$
though, actually, the dependence on $\beta$ could crucially change
the asymptotic behavior: for instance, at $\beta\sim \exp[N]$ one
gets $r\sim x^2\sim \exp[-N]$, and $z^2\sim u^2\sim 1$. The real
situation is clarified after the appropriate analysis of the
experimental data set.

\section{Comparing Data with Experiment\label{III}}

As we have mentioned in the Introduction, a complete analysis of
inflation models versus the experimental situation can be found in
\cite{DdeVegaS1,DdeVegaS2}, having presented the evaluation of
parameters in the framework of Monte Carlo Markov Chains under the
slow-roll approximation of theoretical entries.

\begin{figure}[htbp]
%\setlength{\unitlength}{1mm}
%\centering{\includegraphics[width=160\unitlength]{ori.eps}}
\begin{center}
%\begin{picture}(120,76)
% \put(0,0){\includegraphics[width=120\unitlength]{ori.eps}}
% \put(110,0){$n_S$}
% \put(2,70){$r$}
%\end{picture}
\setlength{\unitlength}{1mm}
\begin{picture}(140,70)
\put(5,3){\includegraphics[width=60\unitlength]{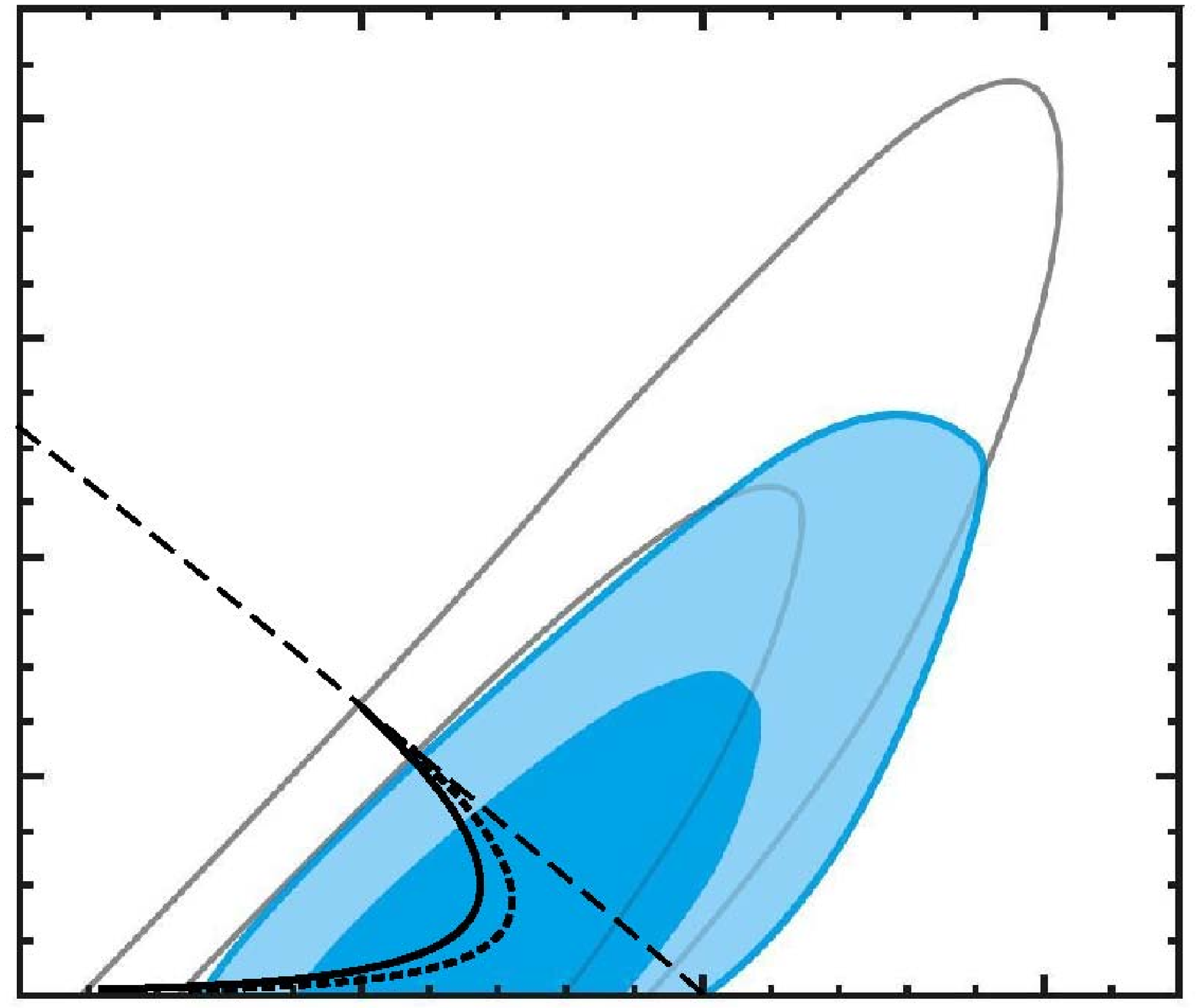}}
\put(76,2){\includegraphics[width=62\unitlength]{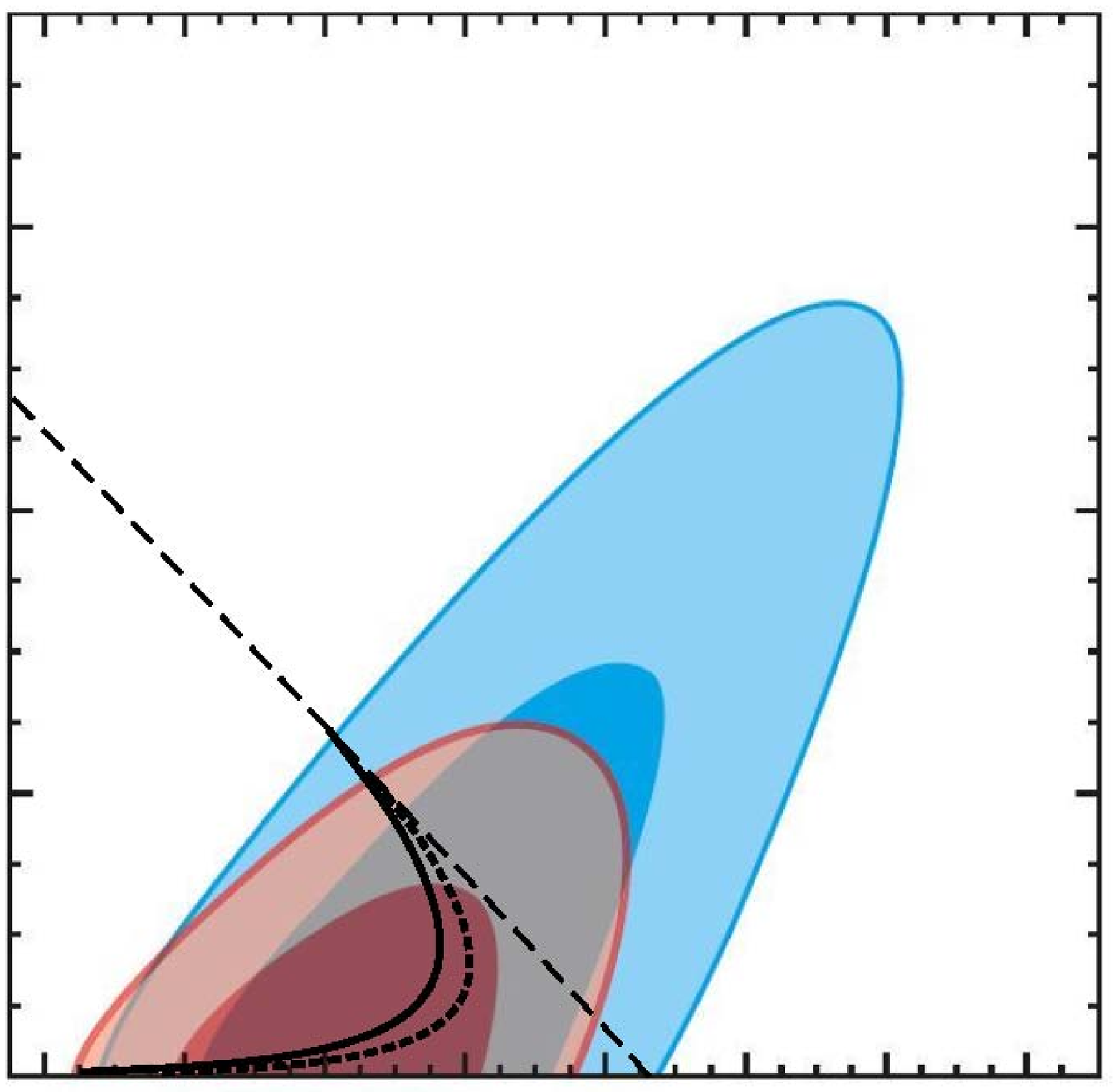}}
 \put(5,1){$0.90$}
 \put(20,1){$0.95$}
 \put(37,1){$1.00$}
 \put(54,1){$1.05$}
 \put(62,1){$n_s$}
 \put(0,15.3){$0.2$}
 \put(0,26){$0.4$}
 \put(0,36.5){$0.6$}
 \put(0,47){$0.8$}
 \put(2,53){$r$}
%%%%%%%%%%%%%%%%%%%%%%
 \put(80,1){$0.92$}
 \put(93,1){$0.96$}
 \put(106,1){$1.00$}
 \put(120,1){$1.04$}
 \put(130.5,1){$n_s$}
 \put(75,17.4){$0.2$}
 \put(75,30.5){$0.4$}
 \put(75,43){$0.6$}
 \put(77,53){$r$}
\end{picture}
\end{center}
\caption{Data of the WMAP collaboration in the plane of the spectral
parameter and the fraction of the tensor term in fluctuations of
density: $\{n_S,r\}$, in comparison with theoretical predictions at
different values of e-folding $N=60$ (thick solid line) and $N=70$
(dotted line), corresponding to the exit of the fluctuation from the
event horizon before the end of inflation (see the text). The left
panel gives contours representing the WMAP data after 3 years of
data taking the confidence levels equal to 1-$\sigma$ and
2-$\sigma$, while the shaded regions give the same confidence levels
after 5 years of data sampling. The right panel shows the WMAP data
after 5 years of data taking in comparison with further constraints
following from BAO and SN experiments.}\label{fig}
\end{figure}

In the present paper, we will use the function in (\ref{N_end}) for
constructing the implicit dependence of $n_S$ versus $r$ at fixed
$N$. Theoretical curves are shown in Fig. \ref{fig} in the
$\{n_S,r\}$-plane (the thick solid line corresponds to $N=60$, the
dotted line does $N=70$). The dashed line shows $u^2=0$, and the
region below it corresponds to the actual case of $u^2>0$, while the
region above it marks $u^2<0$, irrelevant to the present work.

The experimental results obtained by the WMAP collaboration after 3
and 5 years of data taking and published in \cite{WMAP} and
\cite{WMAP5-1,WMAP5-2}, respectively, are presented in Fig.
\ref{fig}, too. The dark shaded contour gives the region with the
1-$\sigma$ confidence level, while the shaded contour corresponds to
the 2-$\sigma$ level. One can see, that the theoretical calculations
are in a good agreement with the experiment at the appropriate
choice of parameters.

From the analysis of data we can obtain quite wide limits of
possible values for parameters of the model potential, namely
\begin{eqnarray}
N=60^{+40}_{-20},\qquad 25\leqslant u^2 \leqslant \infty,
\end{eqnarray}
at the 1-$\sigma$ level, and
\begin{eqnarray}
N=60^{+80}_{-27},\qquad 17\leqslant u^2\leqslant \infty,
\end{eqnarray}
at the 2-$\sigma$ level in the 3 year data sample by WMAP. The
formally infinite vacuum expectation value for the inflaton
certainly corresponds to the final value of its mass, as we shall
see below. The data acquisition of 5 year sample leads to more
strict constraints. So, the above estimates on $N$ with the
confidence level of 1-$\sigma$ transfer to the 2-$\sigma$ level, as
is clearly seen from the figure.

However, the amount of e-folding is, in fact, limited by the actual
history of the Universe evolution after inflation \cite{LL}, so that
the analysis leads to the typical value of $N\approx 60$ (see also
\cite{KT3}). In addition, one has to take into account data of other
experiments: that on baryonic acoustic oscillations and spacial
distribution of galaxies (BAO) \cite{BAO} as well as on the
supernovae Ia (SN)
\cite{Riess:2004nr,Riess:2006fw,Astier:2005qq,WoodVasey:2007jb}.
Such an analysis has been done in \cite{WMAP5-2}, and is presented
in the right panel of Fig. \ref{fig}.

Then, the data at the 1-$\sigma$ level give the constraint on the
parameter $\beta$ in (\ref{r}) and (\ref{ns}) in the form
\begin{equation}\label{beta}
    0.75 \leqslant\beta\leqslant 140.
\end{equation}
The region of $\beta \leqslant 1$ corresponds to the scenario of
chaotic inflation, when the field evolves towards the minimum of
potential from large values at the branch of the potential
approaching infinity, while $\beta >1 $ describes the scenario of
new inflation, when the field ``rolls down'' to the minimum from
small values near the peak at $\phi=0$ (one refers to the case of
``hilltop'' inflation). Indeed, the condition for the critical point
(\ref{16}) during inflation at $y^4 \to 1$ can be approximately
written down in the form
$$
y^2 \approx \frac{9x^2}{4z^2} - u^2z^2 = \frac{3}{16}\,r(1-\beta)
$$
at $\beta < 1$. So, since $y^2 \sim \phi^2-v^2$ one can
straightforwardly see that $\beta = 1$ just separates the regions of
parameters for new and chaotic inflation.

Now let us determine the coupling constant $\lambda$. The WMAP, BAO and
SN observations give
\begin{equation}
P_S=2.457^{+0.092}_{-0.093} \cdot 10^{-9},
\end{equation}
while
\begin{equation}
\lambda=8\pi^2x_c^2z^4P_S=\frac{3\pi^2}{2^{13}}\,r\,
{\big[4r-(r-16)(n_S-1)\big]^2}\,P_S.
\end{equation}
Therefore, at $N\approx 60$ with (\ref{beta}) we get
\begin{equation}
0\leqslant \lambda\leqslant 9.7\cdot 10^{-14},
\end{equation}
while the maximum is located at $\beta\approx 35$. The scale of quartic
coupling is quite natural, if one takes into account the analysis of
$1/N$-expansion during the inflation as performed in
\cite{BdeVegaS1,BdeVegaS2} and mentioned in Introduction.

It is worth noting, that the product of $\lambda u^2$ remains finite
\begin{equation}
\lambda u^2=\frac{\pi^2}{2^{7}}\,r\,
{\big[-3r+(n_S-1)(r-16)\big]}\,P_S.
\end{equation}
Moreover, under (\ref{beta}) the square of inflaton mass in
vicinity of potential minimum
$$
    m^2=\frac{3}{2\pi G}\,\lambda\, u^2
$$
takes the values
\begin{equation}\label{mass}
    1.03\cdot 10^{13}\mbox{ GeV}\leqslant m\leqslant 1.74\cdot
    10^{13}\mbox{ GeV,}
\end{equation}
maximal at $\beta\approx 7$. Furthermore, since
$\mbox{sign}(m^2)=\mbox{sign}(v^2)=\mbox{sign}(u^2)$, the border of
the applicability region for the potential is given by the following
equation
\begin{equation}
n_S-1=-\frac{3r}{16-r},
\end{equation}
which is represented by the dashed line in Fig. \ref{fig}.

Experimental constraints for dependence of the spectral index on the
number of e-folding $N$ in terms of the parameter $d n_S/d\ln N$ are
not restrictive, since they give a value compatible with zero at the
confidence level of 2-$\sigma$, with the quite large uncertainty
being greater than the expected value of this parameter in the model
under study. Therefore, we do not incorporate it into our estimates.

Thus, we see, that one could extract the mass of the inflaton
corresponding to maximal definiteness for all of the potential
parameters.

\section{Conclusion}

Thus, in the present paper we have carried out the analysis of an
inflation model with the inflaton potential including both quadratic
and quartic terms of self-action. The model has allowed us to
consider scenarios of chaotic and new inflation in the framework of
the quasiattractor method, which has enabled us to quite elegantly
calculate the recently observed inhomogeneity of the cosmic
microwave background and distribution of matter in the Universe. We
have shown that such a model is consistent with the observational
data.  One can see, of course, that this model of the potential
\textit{parametrically} cannot satisfy all of the experimentally
admissible values of $n_S$ and $r$ within the empirical
uncertainties (such a potential would not explain the presence of
experimental points above the dashed curve in Fig. \ref{fig}), but
these restrictions are not critical within the accuracy of
measurements, and the given potential seems to be consistent with
the current data.

We have obtained also, that observational data on the inhomogeneity
of the Universe corresponds to the time of forming the inflaton
fluctuations, when the Universe expands approximately $e^{60}$ times
to the end of inflation, which is in agreement with other
estimations. We have also precisely enough determined the inflaton
mass.

The work of V.V.K. is partially supported by the Russian
Foundation for Basic Research, grant 07-02-00417.

\end{document}